\documentstyle[aps,harvard,epsfig,multicol]{revtex}      % Letter

\begin{document}

%----------------------------------------------------------------------
%                       NEW COMMANDS
%----------------------------------------------------------------------
\newcommand{\mycaption}[2]{\begin{center}{\bf Figure \thefigure}\\{#1}\\{\em #2}\end{center}\addtocounter{figure}{1}}
\newcommand{\myauthor}{Ingve Simonsen and Kim Sneppen}
\newcommand{\mytitle}{Profit Profiles in Correlated Markets}
\newcommand{\r}[1]{(\ref{#1})}

%----------------------------------------------------------------------
%                               TITLE AND AUTHORS
%----------------------------------------------------------------------
\title{\mytitle} 
\author{        Ingve Simonsen\footnote{Fax: +45 +45 35 38 91 57}    
                                %\footnote{Email : ingves@nordita.dk} 
                 and
                Kim Sneppen%\footnote{Email : sneppen@nordita.dk}
       }

\address{
          Nordic Intitute for Theoretical Physics --- NORDITA\\
          Blegdamsvej 17, DK-2100 Copenhagen {\O}\\
          DENMARK
        }

\date{\today}
\maketitle

%----------------------------------------------------------------------
%                                       ABSTRACT 
%----------------------------------------------------------------------
\begin{abstract}
    We consider a financial market where the asset price follows a
    fractional Brownian motion. We introduce a family of investment
    strategies, and quantify profit possibilities for both persistent
    and antipersistant markets.
\end{abstract}

\pacs{Key words: Hurst exponent, correlations, non equilibrium, profit, strategies}
\pacs{JEL classification: C5 (Econometric Modeling), 
          D5 (General Equilibrium and Disequilibrium)}
\draft

% --------------------------------------------------------------------
%                                       MAIN TEXT
%---------------------------------------------------------------------

Financial time series have been studied for a long time. What type of
random process they resemble is still debated, and different
suggestions have been made; In 1900 the French mathematician L.\ 
Bachelier~\cite{Bachelier1900} suggested that asset prices might be
described by what today is known as a {\em random
    walk}~\cite{Book:Feder}. Such a walk is a random process where the
increments are {\em uncorrelated}.  However, today we know this is not
the whole story. There are examples of financial time-series that
exhibits correlations
~\cite{Book:Mantegna,Book:Bouchaud,Book:Peters,Peters1996,Mulligan}.
In particular both correlations~\cite{Peters1996,Mulligan} and
recently also anti-correlations~\cite{Mulligan} have been reported for
various types of financial time series.

Correlated markets are at odds with an efficient market, as it may
allow for arbitrage opportunities.  This should be the case both for
short term correlations, as well as long term scale invariant
fluctuations (where transaction costs can be neglected).  This paper
address the amount of arbitrage as function of correlations, and thus
quantify the degree of non-equilibrium and possible profits, for
different employed strategies, a given long time correlation
corresponds to.  For a record of real markets with significant
correlations, please consult~\cite{Mulligan,Neely1997,Holst2000} (and
references therein) \medskip

The price model that we are going to consider is a {\em fractional
    Brownian motion}, {\it i.e.} a self-affine
process~\cite{Book:Feder}.  Such a process is characterized by an
exponent, $H$, termed the Hurst exponent~\cite{Hurst1951}, where
typically $0<H<1$.

Let $p(t)$ denotes the price of an asset at time $t$. The asset could
be a stock, a currency or some commodity.  That $p(t)$ is self-affine
means that fluctuations on different timescales can be rescaled, so
that the behaviour of $p(t)$ is statistically equivalent to the
behaviour of $\lambda^{-H} p(\lambda t)$ where $\lambda$ is any
positive number.  For $H=1/2$ we have the uncorrelated random walk of
Bachelier.

Since it will be useful later, let $t=0$ be the present time, and let
us consider the correlation function between future and past price
increments $\Delta p(t) = p(t) - p(0)$ and $\Delta p(-t)$:
\begin{mathletters} 
    \label{Eq:Corr-func}
\begin{eqnarray}
  \label{Eq:Corr-func-A}
  C_H(t) &=& \frac{
                    \left< -\Delta p(-t) \Delta p(t) \right>
                 }{
                    \left<\left[\Delta p(t)\right]^2\right>
                  },
\end{eqnarray}
where $\langle .. \rangle= \frac{1}{N} \sum_{1}^{N} ..$ denotes the
(arithmetic) average over all considered investment situations for
fixed $H$ (and investment strategy), and thus $\langle \Delta p(t)^2
\rangle =\langle \Delta p(-t)^2 \rangle$.  A remarkable feature of a
fractional Brownian motion is that the correlation function,
Eq.~\r{Eq:Corr-func-A}, is time-independent and only depends on the
Hurst exponent $H$~\cite{Book:Feder}:
\begin{eqnarray}
  \label{Eq:Corr-func-B}
  C_H(t) &=&  2^{2H-1}-1.
\end{eqnarray}
\end{mathletters} 
Thus an ordinary Brownian motion with $H=1/2$ have $C_{H=1/2}(t)=0$
whereas any larger (smaller) $H$ implies correlation
(anti-correlation).  {\it I.e.} if $H>1/2$ then $C_H>0$ and thus the
sign of the past price difference $-\Delta p(-t)=p(0)-p(-t)$ is most
likely to be maintained for $\Delta p(t)$.  Therefore for $H>1/2$ the
stochastic process is most likely to keep the trend of the past.
Similarly if $H<1/2$ the sign of the price difference is most likely
to change at each subsequent time interval.  Furthermore, the
propability of change is independent of the length of the considered
time interval.  Hence, a process with $H>1/2$ is denoted a {\em
    persistent} fractional Brownian motion, while a process with
$H<1/2$ is referred to as {\em anti-persistent}. In the financial
literature the latter case ($H<1/2$) is often referred to as a {\em
    mean-reverting} process.  
\smallskip

To study profit opportunities in a fractional Brownian market, we need
to define a strategy for when to (and not to) make an investment.  In
accordance with the notion of persistence and anti-persistence, we use
a strategy that is based on the price history: Assume that we have an
investment horizon $T$, meaning that after purchasing an asset at time
$t=0$ the investor sticks to his investment until time $t=T$ at which
point the asset is sold.  We will evaluate strategies, assuming that
there is only one option to buy or not to buy at the beginning of each
interval.  If we do not buy, nothing happens. If, however, one buys,
one is bound to sell at the end of the interval. Buying or not buying
at the next interval, is considered a separate independent event.

The choice of strategy, $\Pi_\nu$, is motivated by
Eqs.~\r{Eq:Corr-func}.  For an investment horizon $T$ we make a choice
at $t=0$ based on the asset price at $t=-T$: If the market is
persistent ($H>1/2$), one invests at $t=0$ only if the price went up
from $t=-T$ to $t=0$, {\it i.e.} if $-\Delta p(-T)=p(0)-p(-T)>0$.
Similarly if the market is anti-persistent ($H<1/2$) one invests at
time $t=0$ only if $-\Delta p(-T) < 0$, {\it i.e.} if the price over
the chosen horizon $T$ (in the past) has decreased.  The above
strategy (obtained by setting $\nu=0$ in the formulae below) can be
expressed in terms of the amount invested at time $t=0$:
\begin{eqnarray}
    \label{Eq:Strategy}
    \Pi_\nu (\Delta p(-T)) &=& \theta\left( - ( H - 1/2 )
                     \; \Delta p(-T) \right)  \; \cdot \;
                \frac{ \left| \Delta p(-T)  \right|^\nu }
                     {\langle |\Delta p(-T) |^{\nu} \rangle }
\end{eqnarray} 
where $\theta(x)=1$ for $x \geq 0$, and $\theta(x)=0$ for $x <0$.  We
will later return to other choices for the parameter $\nu$.  In
Eq.~\r{Eq:Strategy} the $\theta$-function signifies to invest
($\theta=1$), or not-invest~($\theta=0$), while the last term define
the investment size for a particular past.  If we always invest the
same amount independent of the past, we have $\nu=0$ so that the last
term of Eq.~\r{Eq:Strategy} is constant (and equal to one).  If $\nu >
0 $ the size of the investment depends on the past change of $p(t)$.
For example if $\nu=1$ the amount we invest is proportional to $\Delta
p(-T)$ whereas higher $\nu$ values focus transactions on fewer big
investments.  \smallskip

\begin{figure}[htbp]
  \begin{center}
    \leavevmode
    \epsfig{file=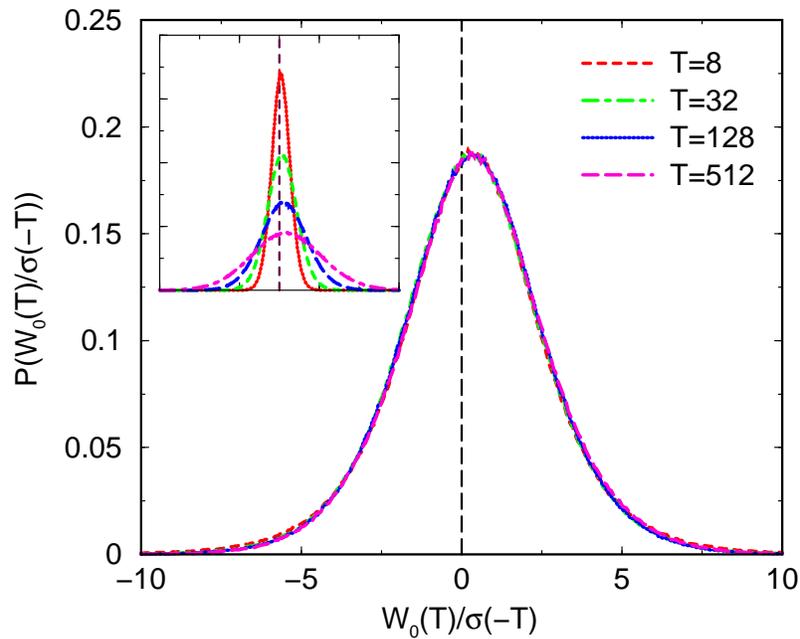,width=10.5cm}
    \caption{The profit distribution for an anti-persistent market
        ($H=0.3$) with different investment horizons $T$ (and fix
        strategy $\nu=0$).  The figure shows the distribution of
        $W_0(T)/\sigma(-T)$ while the inset shows the distribution of
        $W_0(T)$.}
    \label{Fig:Wealth03}
  \end{center}
\end{figure}

To measure the utility of a strategy $\Pi_{\nu}(\Delta p(-T))$ we note
that the profit for an investment (occuring only when $\theta=1$ in
Eq. \r{Eq:Strategy}) is
\begin{eqnarray}
    \label{Eq:Wealth}
    W_\nu(T) &=&  \frac{ \left| \Delta p(-T)  \right|^\nu }
                 {\langle |\Delta p(-T) |^{\nu} \rangle } \; 
                 \cdot \; \Delta p(T).
\end{eqnarray}
If the price fluctuations are increased, the potential profit (or
loss) is bigger. In our fractional Brownian market the fluctuations
increase with the duration of the investment horizon. Therefore the
profit histograms are expected to broaden when $T$ is increased (see
inset in Fig.~\ref{Fig:Wealth03}). In order to compare the performance
of the various strategies at different horizons we examine profits in
units of the typically price fluctuations at the given investment
horizon: $\sigma(-T)=\sqrt{ \langle (p(-T)-p(0))^2 \rangle } \propto
T^H$ for a stochastic process with Hurst exponent $H$.  Hence it is
natural to consider the probability distribution of
$W_\nu(T)/\sigma(-T)$.

\begin{figure}[htbp]
  \begin{center}
    \leavevmode
    \epsfig{file=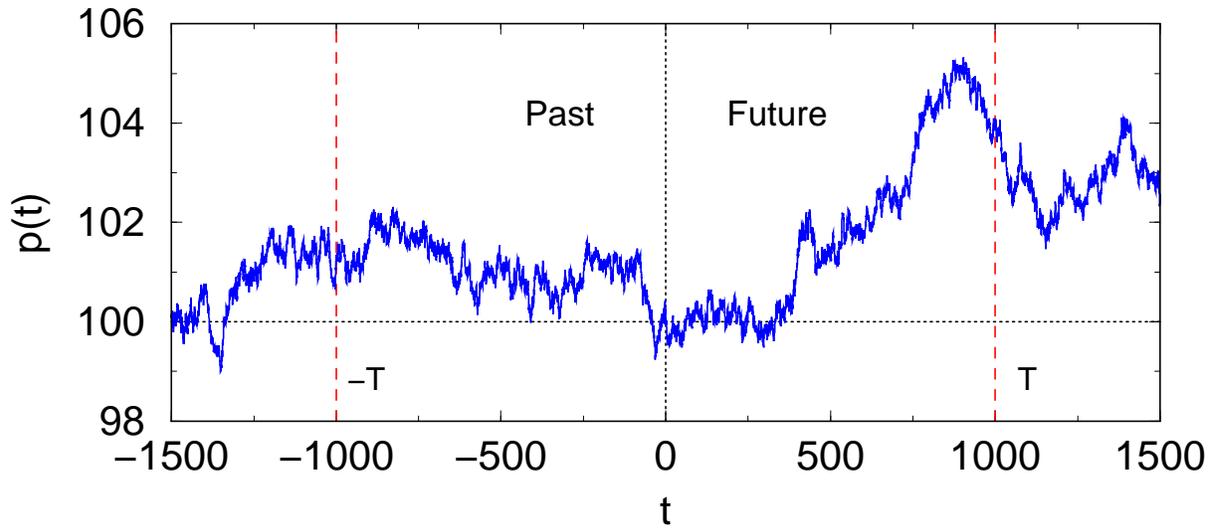,width=16cm}    
    \caption{An example of a price time series with Hurst exponent 
        $H=0.4$.  With the strategy described in the text, one would
        for time horizon $T=1000$, buy at $t=0$ because the price went
        down in the past and thus is more likely to increase in the
        future.}
    \label{Fig:Prices}
  \end{center}
\end{figure}

To obtain the distribution of profit we need to generate a market.
Fractional Brownian price processes were generated by the method
described in Ref.~\cite{Book:Feder} which ensures Gaussian increments
of the process. For $H=0.4$ an example is shown in
Fig.~\ref{Fig:Prices}. We have also considered versions of the
generator described in Ref.~\cite{Book:Feder} which gives raise to
non-Gaussian increments, {\it e.g.} power-laws~\cite{power-law}, and
different generators~\cite{Prakash,WG}, but these different choices
seem not to affect the general conclusions drawn in this paper.  The
length of the used time series was $T_{max}=2^{14}=16384$, equally
divided between past and future.  To avoid influence from periodic
boundaries we only consider horizons $T$ well below $T_{max}/2$.

\begin{figure}[htbp]
  \begin{center}
    \leavevmode
    \epsfig{file=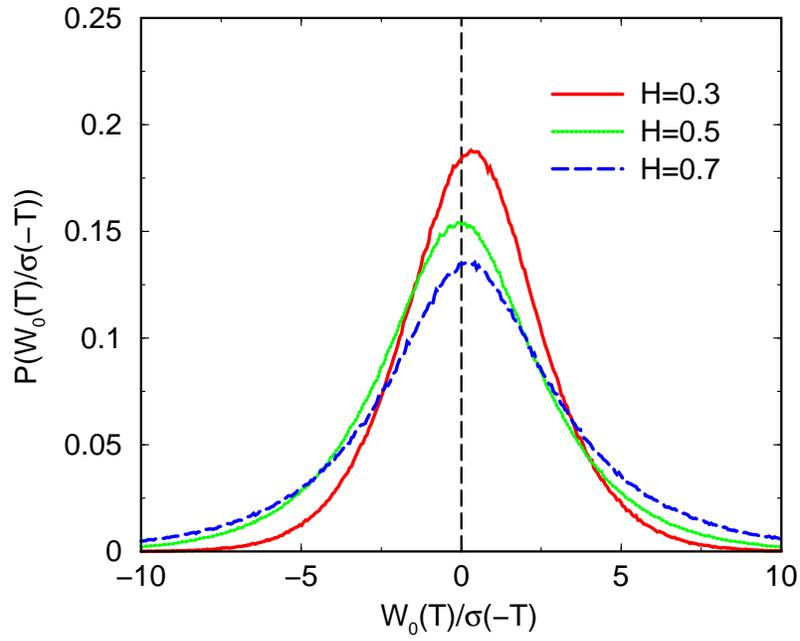,width=10.5cm}
    \caption{The distributions of $W_0(T)/\sigma(-T)$ 
        for markets with different Hurst exponents $H$ (and investment
        strategy $\nu=0$).}
    \label{Fig:Several}
  \end{center}
\end{figure}

In the inset of Fig.~\ref{Fig:Wealth03} we show, for an
anti-persistent market characterized by Hurst exponent $H=0.3$ (and
strategy $\nu=0$), the profit probability distributions for the
investment horizons $T=8,32,128,512$. The corresponding distributions
for the renormalized profits, $W_0(T)/\sigma(-T)$, still with $H=0.3$,
are shown in the main part of Fig.~\ref{Fig:Wealth03}.  They
demonstrate data collapse for the different investment horizons.  This
teaches us that any time-scale provides the same profit profile, if we
measure this in units of the typical price fluctuations on that
time-scale (and can neglect transaction costs). This fact simplifies
our discussion, as we can deal with profit in terms of only fractions
of the typical price fluctuations.  The vertical dashed lines in
Fig.~\ref{Fig:Wealth03} separate the profit-region (to the right) from
the loss-region (to the left).  The asymmetry testifies to a possible
mean profit, which we for Hurst exponent $H=0.3$ find to be
$\left<W_0(T)/\sigma(-T)\right>_P\;=\;0.35$ where
$\left<\cdot\right>_P$ is used to indicate the mean relative to the
distribution $P$ (see Ref.~\cite{Distribution}).  We find similar
data-collapses for other Hurst exponents between $0$ and $1$.
Furthermore, in Fig.~\ref{Fig:Several} the collapsed curves, for
different $H$, show that the profit distribution exhibits fatter tails
for larger Hurst exponents.  We also observe that only $H=1/2$ gives
raise to a symmetric distribution, whereas all other markets allow for
profit.

\begin{figure}[htbp] 
  \begin{center}
    \leavevmode
    \epsfig{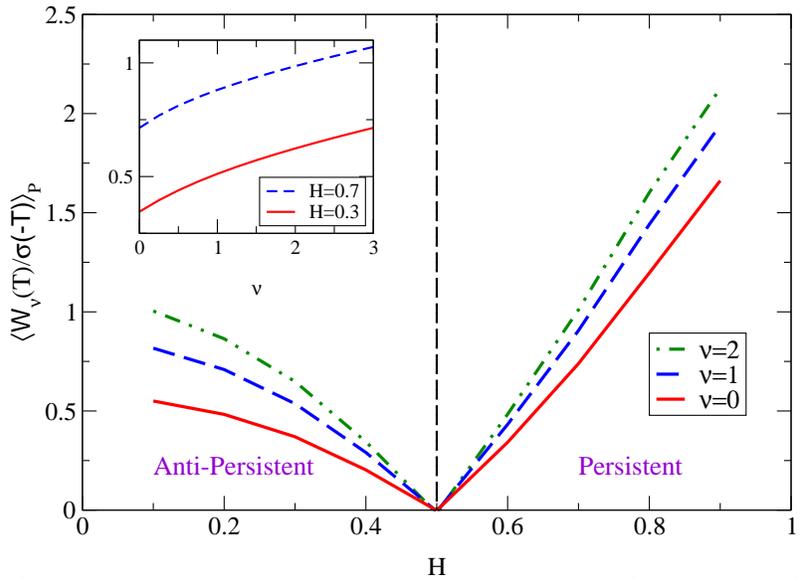}
    \caption{The average profit
        $\left<W_\nu(T)/\sigma(-T)\right>_P$ vs.  Hurst exponent, $H$,
        for three different choices of strategy. The inset shows the
        same quantity for $H=0.3$ and $0.7$, but now as a function of
        the invetsment strategy $\nu$.}
    \label{Fig:Total} 
  \end{center}
\end{figure}

The above analysis was done for a $\nu=0$ strategy, {\it i.e.}  many
equally sized investments.  If one instead choose to focus the
investments, that means to use a $\nu>0$ strategy, the profit would be
different (see Fig.~\ref{Fig:Total}). However, the
distributions~\cite{Distribution} of $W_\nu(T)/\sigma(-T)$ (results
not shown), for different $\nu$ and given $H$, are rather similar.
The main difference between the different strategies $\nu$ comes about
due to an increased asymmetries (for $H\neq 1/2$).  This asymmetry is
quantified by the average profit $\left<W_0(T)/\sigma(-T)\right>_P$.
In Fig.~\ref{Fig:Total} this average profit is summarized for a number
of markets and a number of different strategies.  For any $\nu$ we
observe profit possibilities for both $H < 1/2$ and $H > 1/2$. For
small deviations from a perfect Brownian market the profit increases
linearly with the distance to $H=1/2$.  By examining different $\nu$
we further observe that larger $\nu$ provides us with larger average
profit.  Thus, if the average is our only concern we should invest
only when there has been extreme price changes in recent past.
However, the cost of higher $\nu$ is a higher probability for big
losses (and gains) simply because the invested capital increases with
$\nu$. This fact was observed (result not shown) as fatter and fatter
tails for the {\em total} wealth distribution. Notice, however, that
the probability for big losses {\em per invested dollar} is more or
less independent of $\nu$.

Finally we consider the role of transaction costs.  Until now this has
been neglected, as it could be for a long time investment. However,
since the profit defined above always scales with the spread in
prices, $W(T)\propto \sigma (T) \propto T^H$, the profit per time unit
$W(T)/T\propto T^{H-1}$ favors (if we neglect transaction costs) very
short investments (since $H<1$).  The only limit to short term
investments is transaction costs, that is a fixed fraction $r$ of
every investment.  With transaction costs the net profit for an
investment $W_\nu(r,T) \propto ( \Delta p(T) - r )$ where $\Delta
p(T)$ here is defined as the relative price change in order to be on
the same scale as the transaction cost $r$.  In principle there is of
course transaction costs at both beginning and end of each trade, but
that would not change the functional form ({\it i.e.}  $\Delta p -r
\rightarrow (1-r) \Delta p-2r$). With transaction costs the average
$W_\nu (r,T)$ is negative at small $T$ and $\langle W_\nu(T)\rangle
/T$ has a maximum at some finite time $T$ that depends on Hurst
exponent, as well as $\nu$ and $r$.  This maximum sets the optimal
investment horizon in a correlated market.  \smallskip

In summary, we have considered the possibility of making profit in a
fractional Brownian market.  It is noted that one can make profit for
all cases where $H\neq 0.5$, and it is found that the average profit
increases with the willingness to bet on extreme variations.

%\acknowledgements
\medskip 

We would like to thank P. Bak for initially drawing our attention to
the problems addressed in this letter.

%%%%%%%%%%%%%%%%%%%%%%%%%%%%%%%%%%%%%%%%%%%%%%%%%%%%%%%%%%%%%%%%%%%%%%
% --------------------------------------------------------------------
%                               BIBLIOGRAPHY
% --------------------------------------------------------------------
%%%%%%%%%%%%%%%%%%%%%%%%%%%%%%%%%%%%%%%%%%%%%%%%%%%%%%%%%%%%%%%%%%%%%%
\bibliographystyle{agsm}  
\bibliography{paper2000-5-bibtex}     

%%%%%%%%%%%%%%%%%%%%%%%%%%%%%%%%%%%%%%%%%%%%%%%%%%%%%%%%%%%%%%%%%%%%%%
% --------------------------------------------------------------------
%                               FIGURE CAPTIONS AND FIGURES
% --------------------------------------------------------------------
%%%%%%%%%%%%%%%%%%%%%%%%%%%%%%%%%%%%%%%%%%%%%%%%%%%%%%%%%%%%%%%%%%%%%%
\widetext
\newpage

\vspace{1cm}

\end{document}